\documentclass[runningheads]{llncs}
\usepackage[T1]{fontenc}
\usepackage{graphicx}

\usepackage[hidelinks]{hyperref}

\begin{document}

\title{OAEI Machine Learning Dataset for Online Model Generation}

\author{Sven Hertling\inst{1}\orcidID{0000-0003-0333-5888} \and 
Ebrahim Norouzi\inst{1}\orcidID{0000-0002-2691-6995} \and \\ 
Harald Sack\inst{1,2}\orcidID{0000-0001-7069-9804}}

\institute{FIZ Karlsruhe – Leibniz Institute for Information Infrastructure, Hermann-von-Helmholtz-Platz 1,
76344 Eggenstein-Leopoldshafen, Germany\\
\email{firstname.lastname@fiz-karlsruhe.de}\\
\and
Karlsruhe Institute of Technology (AIFB), Kaiserstr. 89, 76133 Karlsruhe, Germany\\
\email{firstname.lastname@partner.kit.edu}
}

\maketitle

\begin{abstract}

Ontology and knowledge graph matching systems are evaluated annually by the Ontology Alignment Evaluation Initiative (OAEI). More and more systems use machine learning-based approaches, including large language models.
The training and validation datasets are usually determined by the system developer and often a subset of the reference alignments are used. This sampling is against the OAEI rules and makes a fair comparison impossible. Furthermore, those models are trained offline (a trained and optimized model is packaged into the matcher) and therefore the systems are specifically trained for those tasks.
In this paper, we introduce a dataset that contains training, validation, and test sets for most of the OAEI tracks.
Thus, online model learning (the systems must adapt to the given input alignment without human intervention) is made possible to enable a fair comparison for ML-based systems. We showcase the usefulness of the dataset by fine-tuning the confidence thresholds of popular systems.

\end{abstract}

\section{Introduction}
If applications want to use two or more knowledge graphs (KGs) simultaneously, the corresponding ontologies and instances must be aligned.
This process is called ontology alignment (more generally, KG alignment). The inputs are two KGs ($KG_{1}$ and $KG_{2}$) as well as an input alignment $A_{in}$. The produced result is an improved alignment $A_{out}$. Each alignment consists of (possibly) multiple correspondences in the form $<e_{1}, e_{2}, r, c>$ where $e_{1} \in KG_{1}$ and $e_{2} \in KG_{2}$. $r$ represents the relation between the entities, such as equivalence or subsumption relation (in case of class correspondences), and $c \in [0,1]$ is a confidence value given by the matching process.

Starting from 2004, the Ontology Alignment Evaluation Initiative (OAEI) evaluates matching systems each year and allows for a fair comparison between them.
Over the years, more and more systems required correspondences for training and validating their machine learning-based approaches.
Especially in 2023, systems for the conference track heavily relied on some form of training data\footnote{\url{https://oaei.ontologymatching.org/2023/results/conference/index.html}} (due to the low number of correspondences in the reference alignment, this makes a huge difference in the final result metrics) but also in other tracks those systems would like to adapt their matching behavior to the task at hand.

In this paper, we introduce train, validation, and test splits for common OAEI datasets and argue that the models need to adapt to the given input alignment during the execution of the systems (online model generation) instead of downloading and generating a model by hand and package it into a matching system (offline model generation). With the latter, the systems will hardly be applicable to new datasets, which is one huge advantage of all matching systems participating at OAEI.

\section{Related Work}

Many systems submitted to OAEI require training data, and in the following, we list those systems together with their setup for generating training examples.
In 2023, \emph{GraphMatcher}~\cite{DBLP:conf/om2/Efeoglu23} used the reference alignment in its 5-fold cross-validation 
and \emph{TO\-MA\-TO}~\cite{DBLP:conf/om2/RoussilleT23} splitted the created dataset into 60\% for training and 40\% for testing.
SEBMatcher~\cite{DBLP:conf/semweb/GosselinZ22} in 2022 generated a training dataset by reference and string alignments and sampled 20 \% of the reference alignment to create positive cases.
Fine-TOM~\cite{DBLP:conf/semweb/KnorrP21} (participating in OAEI 2021) used a transformer architecture that needed fine-tuning. Their dataset included 20\% of each reference alignment from the Anatomy, Conference, and Knowledge Graph track.

One first direction in creating a machine learning dataset for OAEI is done by He et al.~\cite{he2022machine}.
They provide a train, validation, and test split for a new track called Bio-ML.
In their evaluation, they used systems that are trained offline, which means the developers download the training alignments, tune the model, and upload a matcher including that model.

\section{Generation of the dataset}

We use the defined tracks of OAEI and split the reference alignment into 20\% for training, 10\% for validation, and 70\% for testing to align with He et al.~\cite{he2022machine} and most other systems that generate training data. The reason why the training and validation fraction is so low is the extreme imbalance between correct and incorrect correspondences. Furthermore, in the real-world setting, only a few positive correspondences will be provided as a training signal due to the high effort of creating those correspondences.

To have a good segmentation of the reference alignment, we stratify it by the following criteria:
(1) entity type (class, property, instance) of the source and target entity, (2) relation type, and (3) difficulty of the mapping.
For the latter, we further differentiate between easy, medium, and hard matches. Easy matches can be found by simple label matching with a bit of string processing like lowercasing and camel case splitting. 
Medium matches share at least one token among the labels, and the rest are considered hard matches.

The stratification for all criteria is achieved by creating subgroups for each combination, e.g., the group class-class-equivalence-easy and instance-instance-equivalence-medium are two subgroups that we use for stratification, such that in training, validation, and test, we ensure the same distribution as in the original reference alignment. 
We include the easy matches in the training and validation alignment because, in the real-world use case, the provided matches will also contain those simple correspondences (especially if the mapping is created by finding matching entities for a random sample of entities). 
The input to a matching system is only one alignment. To differentiate between the training and validation set, we added an additional correspondence extension to indicate to which set it belongs. If a system ignores this distinction, it can use the whole input as training data.

Only positive training examples are provided as input.
For a machine learning model, negative examples are often required. Unlike \cite{he2022machine}, we do not provide negatives (they need it for easily evaluating a ranking-based method) but let the system create them on the fly.
This can be achieved by assuming that, for an entity in the source, at most one entity exists in the target graph (or vice versa), and both graphs are duplicate-free.
Then, given a correspondence $<A, B>$ in the training alignment, the system can use its own distance function to search for hard negatives of A and B. 
The advantage is that those correspondences are hard negatives for the matching system at hand and reflects the amount and distribution during the matching of the whole input KGs.

We used only OAEI test cases with at least 350 correspondences in the reference alignment, such that the combined training and validation sets have at least 100 correspondences. This results in the following tracks: Anatomy, BioDiv, Knowledge Graph, and Bio-ML. 

All datasets are well integrated with MELT~\cite{hertling2019melt} and can be downloaded via the track repository\footnote{\url{https://dwslab.github.io/melt/track-repository\#ml-based-tracks}} (there is a separated section for it). Code for the generation and evaluation can be found on GitHub\footnote{\url{https://github.com/dwslab/melt/tree/master/examples/mlDataset}}.

\section{Use Case}

We analyze the usefulness of the presented dataset by using three matching systems from OAEI 2023, which return correspondences with a wide range of confidence values (not only 1.0). This results in the following systems: Matcha \cite{DBLP:conf/om2/FariaSCFBP23}, LogMap \cite{DBLP:conf/om2/Jimenez-Ruiz23,jimenez2011logmap}, and OLaLa \cite{DBLP:conf/om2/HertlingP23,DBLP:conf/kcap/HertlingP23}. We use the training and validation data to adapt the confidence threshold to the task at hand and remove correspondences below the threshold, which could improve precision and eventually lower recall.
Two automated approaches for finding the right threshold are implemented under the assumption that the input alignment is partial (supervised partial - SPart) or complete (supervised complete - SComp). It computes the optimal threshold by using the training and validation data as a ground truth and finds the value that gives the highest F$_1$-Measure.

\begin{table}[t]
    \caption{Results of unsupervised (U) and two supervised approaches (SPart, SComp) for tuning the confidence threshold.}
    \centering
	\label{tab:results}
\begin{tabular}{|l||r|r|r||r|r|r||r|r|r|}
\hline
\multicolumn{1}{|l||}{} & \multicolumn{3}{c||}{Precision} & \multicolumn{3}{c||}{Recall} & \multicolumn{3}{c|}{F$_1$-Measure} \\ \hline
                       & U        & SPart       & SComp       & U       & SPart      & SComp      & U        & SPart       & SComp      \\\hline\hline
Anatomy                &          &          &          &         &         &         &          &          &          \\
\hspace{0.2cm} Matcha  & 93.09    & 93.42    & \textbf{94.72}    & \textbf{92.65}   & 92.37   & 91.23   & 92.87    & 92.89    & \textbf{92.94}    \\
\hspace{0.2cm} LogMap  & 88.34    & 89.86    & \textbf{96.85}    & \textbf{85.01}   & 84.35   & 69.46   & 86.65    & \textbf{87.02}    & 80.90    \\
\hspace{0.2cm} OLaLa   & 89.04    & 89.27    & \textbf{95.93}    & \textbf{89.63}   & 89.35   & 82.09   & \textbf{89.34}    & 89.31    & 88.47    \\\hline

BioDiv                 &          &          &          &         &         &         &          &          &          \\
\hspace{0.2cm} Matcha  & 62.30    & 63.32    & \textbf{63.69}    & \textbf{99.12}   & 99.07   & 99.05   & 76.51    & 77.26    & \textbf{77.53}    \\
\hspace{0.2cm} LogMap  & 61.02    & 61.86    & \textbf{62.72}    & \textbf{99.25}   & 98.30   & 93.84   & 75.58    & \textbf{75.93}    & 75.19    \\
\hspace{0.2cm} OLaLa   & 63.37    & 63.69    & \textbf{90.37}    & \textbf{99.27}   & 99.18   & 90.63   & 77.36    & 77.57    & \textbf{90.50}    \\\hline

KnowledgeGraph         &          &          &          &         &         &         &          &          &          \\
\hspace{0.2cm} Matcha  & 7.01     & 13.35    & \textbf{33.09}    & \textbf{80.64}   & 74.88   & 56.63   & 12.90    & 22.66    & \textbf{41.78}    \\
\hspace{0.2cm} LogMap  & 45.86    & 45.90    & \textbf{46.31}    & \textbf{72.65}   & \textbf{72.65}   & 72.56   & 56.23    & 56.26    & \textbf{56.54}     \\ \hline

Bio-ML                 &          &          &          &         &         &         &          &          &          \\
\hspace{0.2cm} Matcha  & 61.37    & 61.45    & \textbf{63.78}    & \textbf{60.29}   & 60.18   & 57.31   & 60.80    & \textbf{60.81}    & 60.37    \\
\hspace{0.2cm} LogMap  & 58.22    & 58.46    & \textbf{64.51}    & \textbf{58.43}   & 58.27   & 53.46   & 58.33    & 58.36    & \textbf{58.47}    \\
\hspace{0.2cm} OLaLa   & 36.42    & 37.31    & \textbf{51.11}    & \textbf{42.30}   & 42.04   & 34.33   & 39.14    & 39.53    & \textbf{41.07}    \\\hline
\end{tabular}
\end{table}

Table~\ref{tab:results} shows the micro-averaged results for all tracks and matchers. The numbers represent the class matches except for the Knowledge Graph track, where the instance matches are shown (OLaLa is not capable of matching instances). The unsupervised case is evaluated on the same test set so that all numbers are comparable.

In most cases, there is only a slight improvement because the systems are highly tuned, and the alignment filtering relies solely on the confidence values.
Nevertheless, in some cases, huge improvements can be observed. For example, when setting the right confidence threshold for OLaLa in the BioDiv dataset, there is a huge improvement of over 13\% in terms of F$_1$-Measure.
We hope to see larger improvements if systems use it to train their whole approach using more features to differentiate between correct and incorrect mappings.

\section{Conclusion}
In this paper, we introduced a new dataset for machine learning systems based on existing OAEI tracks.
With the presented dataset, we hope to encourage matching system developers to use a given input alignment to tune and optimize their parameters online and only fall back to default/pre-trained parameters if no input alignment is given.

In the future, we plan to create datasets for the conference track that do not have many correspondences. Thus, one complete test case can be used as training data and another one as a test (in-domain transfer learning).

\bibliographystyle{splncs04}
\bibliography{main}
\end{document}